\documentclass[english,twocolumn]{emulateapj}

\usepackage[T1]{fontenc}
\usepackage[latin9]{inputenc}
\usepackage{babel}

\usepackage{graphicx}

\makeatletter

\providecommand{\tabularnewline}{\\}

\usepackage{rotating}

\newcommand{\lsim}{\mathrel{\rlap{\lower4pt\hbox{\hskip1pt$\sim$}}
    \raise1pt\hbox{$<$}}}\newcommand{\gsim}{\mathrel{\rlap{\lower4pt\hbox{\hskip1pt$\sim$}}
    \raise1pt\hbox{$>$}}}

\def\in{\rm in}\def\out{\rm out}

\makeatother

\shorttitle{Orbital properties of binary minor planets}
\shortauthors{Naoz, Perets \& Ragozzine}
\makeatother

\begin{document}

\title{ The observed orbital properties of binary minor planets}

\author{Smadar Naoz\altaffilmark{1}$^{,}$\altaffilmark{2}, Hagai B. Perets\altaffilmark{3} and Darin Ragozzine\altaffilmark{3} }

\altaffiltext{1}{ CIERA, Northwestern University, Evanston, IL 60208, USA}
\altaffiltext{2}{Raymond and Beverly Sackler School of Physics and \\
Astronomy, Tel Aviv University, Tel Aviv 69978, Israel}
\altaffiltext{3}{ Harvard-Smithsonian Center for Astrophysics, 60 Garden St.,
Cambridge MA 02138, USA}

\begin{abstract}
Many binary minor planets (BMPs; both binary asteroids and binary
Trans-Neptunians objects; TNOs) are known to exist in the Solar system.
The currently observed orbital and physical properties of BMPs hold
essential information and clues about their origin, their evolution
and the conditions under-which they evolved. 
Here we study the orbital properties of BMPs with currently known mutual orbits
 We find that
BMPs are typically highly inclined relative to their orbit around
the sun, with a distribution consistent with an isotropic distribution.
BMPs not affected by tidal forces are found to have high eccentricities
with non-thermal eccentricity distribution peaking at intermediate
eccentricities (typically $0.4-0.6$). The high inclinations and eccentricities
of the BMPs suggest that BMPs evolved in a dense collisional environment,
in which gravitational encounters in addition to tidal and secular
Kozai affects played an important role in their orbital evolution. 
\end{abstract}

\section{Introduction}

\label{intro} The binary asteroids and binary Trans-Neptunian objects (TNOs) discovered in recent years show a large diversity of orbital properties, showing
 a wide range of eccentricities, inclinations,
separations and mass ratios. Many models have
been suggested for the origin of these binary minor planets (BMPs)
and their orbital configurations \citep{ric+06}. 
An essential component
in constraining theoretical models for the origin and evolution of
BMPs is understanding the distribution of their orbital parameters.
The number of BMPs with known orbital parameters is currently small.
Nevertheless, $29$ BMP systems already have full solutions for their
mutual orbits, including $17$ TNOs and $12$ asteroids (we do not
consider near earth objects which have much shorter lifetimes). 
We study the orbital properties of the 29 main belt and 
transneptunian binaries
(which we refer to corporately as BMPs) to provide clues 
to and constraints on their evolutionary history.
Several reviews have presented the observed separations of BMPs \citep[e.g. ][]{ric+06,nol+08b,wal09}
here we focus on the distributions of eccentricities and inclinations,
not shown before. In addition, we discuss the relations between the
orbital parameters of BMPs, including their observed periods/separations.

In the following we present and discuss the published orbital solutions of BMPs.
 We present the distributions of the orbital parameters
of these BMPs, treating binary TNOs and binary asteroids separately. We also
 discuss the selection biases affecting both samples.
We then briefly study the implications of our findings regarding the
conditions in the early Solar system and the formation and evolutionary
scenarios of BMPs.

\section{The orbital parameters of BMPs }

\subsection{The data}

The mutual orbits of $30$ BMPs have been published in the
literature (see tables 1 and 2), some of them with two degenerate
solutions, and a few published with no indicated inclination. Table
1 shows the physical properties and orbital parameters
of BMPs with known inclinations in the solar system.
In the literature the orbital parameters of BMPs are typically given 
with respect to the ecliptic plane, while the inner orbital parameters 
are given with respect to the equatorial frame of reference.
For our analysis we are interested in the mutual inclinations 
between the BMP orbit around the Sun and the BMP
inner orbit, therefore we used the published data to calculate 
these inclinations for our analysis of the dynamical evolution (see Appendix).

\begin{figure}
\centering \includegraphics[width=84mm]{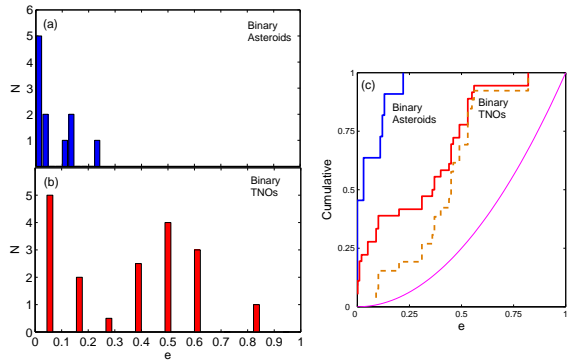} \caption{\label{fig:histe}
Eccentricity distribution of observed binary asteroids
and  TNOs (panels (a) and (b) respectively). Panel (c)  shows the cumulative  
distribution of both samples (asteroids and TNO binaries, solid lines, respectively) as well as comparison to a thermal distribution (lower solid line). The dashed line shows the cumulative eccentricity distribution of binary TNOs excluding binaries that are likely affected by tides ($e<0.05$).}
\end{figure}

\begin{figure}
\centering \includegraphics[width=84mm]{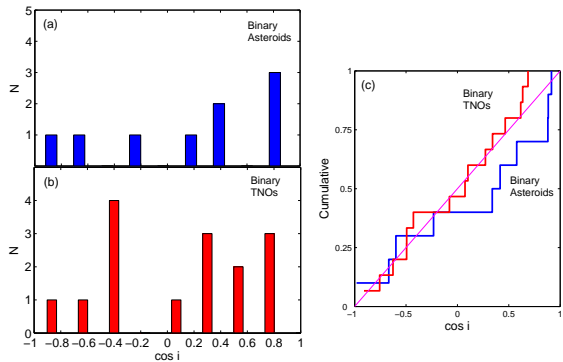} 
\caption{Inclination distribution of observed binary asteroids and binary TNOs  (panels (a) and (b) respectively). Panel (c) shows the cumulative distribution of both samples as well as comparison to a flat distribution (i.e. random in $\cos{i}$). Note that many of the binary TNO inclinations have two degenerate solutions; here only one solution is chosen arbitrarily for each binary to illustrate their likely distribution (see text).}
\label{fig:histi}
\end{figure}

In some cases two degenerate orbital solutions were found for the
BMPs; these solutions differ significantly in their derived inclinations,
but have very similar eccentricities. In these cases we detail both
solutions (see table 1) and we use the published eccentricities to
derive the eccentricity distribution (shown in Fig. \ref{fig:histe}).
All published binary asteroids orbits have unique solution for the
inclination (besides the binary asteroids Balam for which only the
eccentricity is known) and we show their mutual inclination distribution
(in $\cos i$; Fig. \ref{fig:histi}). Only 5 of the binary TNOs, however, have
a unique non-degenerate solutions for their inclinations. We therefore
can not show the true inclination distribution for binary TNOs with
significant statistics (Fig. \ref{fig:histi} shows the distribution
of binary TNOs inclination for an arbitrary choice of one solution
for each binary TNO from the two possible degenerate solutions published).
Nevertheless, we do consider all the possible distributions of the
binary TNOs in a statistical manner (see section 3). 
Note that both binary TNO and binary asteroid populations are presented. Given
that these two populations differed in the conditions under which
they evolved, their orbital properties are presented separately.

\subsection{Selection effects}

There are significant observational selection effects present in the
current distribution of binary TNOs and asteroids. These effects are
very difficult to correct since they depend on several factors, and
the currently observed population of binaries comes from very heterogeneous
observing conditions. Although we do not correct for these effects
in the current analysis, we discuss them below to recognize the
possible biases they may produce. 

\subsubsection{Binary TNOs}

Binary TNOs are primarily discovered in two ways. Ground-based observations
can detect two co-moving objects, if the projected separation of the
objects is roughly larger than the typical seeing, $>$0.5 arcseconds
\citep[e.g. ][]{ker+06} or if adaptive optics is used on bright targets
\citep[e.g. ][]{bro+06}. The widest binaries, such as 2001 QW322 \citep{pet+04} can be found with this mechanism, assuming that discovery and
follow-up images have been searched for binaries (generally true,
but not always). Pan-STARRS and other future surveys will probably
detect dozens of binaries this way \citep{hol+07}. The second discovery
mechanism is using Hubble Space Telescope (HST) observations combined
with PSF-fitting techniques, which has discovered the great majority ($\sim80\,\%$)
of known binary TNOs. HST can resolve binaries as close as tens of
milliarcseconds and generally probes much fainter objects than ground-based
telescopes \citep{nol+08b}.

Each of these two methods suffer from important observational biases. 
The main bias is a detection bias: the secondary must be bright
enough and far enough away from the primary to be detected. In both
the ground- and space-based cases, there is no simple prescription
that describes this detection bias, especially since these are coupled
close to the primary: a bright secondary can be discovered at smaller
separations than a faint secondary. Furthermore, since HST observations
are always nearly the same duration (one HST orbit), observations
are essentially magnitude-limited, implying that low brightness ratios
(and mass ratios) can only be seen around brighter targets.

A discussion of these biases and their effects on the observed population
is also given by \citet{nol+08b}. Despite the observational biases,
these authors believe that there is good evidence that binaries composed
of moderate-sized TNOs (diameters less than $\sim$1000 km) are actually
clustered at nearly equal brightness, with $\Delta_{mag}<1$ corresponding
to a mass ratio of greater than $\sim$0.25 assuming equal albedos
and densities. This latter assumption seems reasonable given that
the colors of components of binary TNOs are known to be similar \citep{ben+09}.

Considering now biases in the binary mutual orbital elements, we again
point out that the majority of binaries are discovered in single HST
snapshots. Hence, biases are introduced by the fact that the orbital
separation must be detectably large at a single epoch. While \citet{nol+08b}
point out that the statistical distribution of observed separations
is similar to the distribution of semi-major axes, it is clear that
systems with larger eccentricities are more likely to be seen since
the observability is increased at apoapse, both due to wider separations
and to the longer residence time. This eccentricity bias is most important
near the angular resolution limit where most of the binaries are discovered,
but it is not important when semi-major axes exceed $\sim$0.2 arcseconds.
Of course, separation-limited observations imply that the smallest
semi-major axes are undetectable; these systems are best discovered
through photometry, either through doubly periodic light curves or
where eclipses and occultations may reveal contact binaries like 2001
QG$_{298}$ \citep{she+04}.

Finally, we also note a bias against observations of  binary TNOs with
large size/mass ratios, due to detection limits.  
For example, a typical TNO with a radius
of 100 km located at 40 AU would have collisional satellites with
$V\,<\,24.5$ at separations of $\sim$0.04 arcseconds, perhaps barely
detectable by HST if the satellite is at elongation. The most likely
way to discover these systems is through photocenter-barycenter shifts
(in this case, the size would be $\sim$2 milliarcseconds) detected
through long-baseline highly-accurate astrometry, potentially available
from future Pan-STARRS or LSST surveys. Detecting a double-periodic
light curve (or serendipitous mutual events) may be possible for some
systems, though the faintness of the components makes this very
difficult and the result may be impossible to distinguish from a single
object with arbitrary shape and spin orientation. The size ratio of
 the binary components can be important for the origin and evolution of 
the binary orbital parameters, and therefore may affect the statistics of
the orbital parameters in our sample.     

\begin{sidewaystable*}
\caption{\protect\footnotesize The orbital parameters of BMPs.}

\begin{tabular}{llllllllllll}
\hline 
{\scriptsize Name}  & {\scriptsize Satellite}  & {\scriptsize Period}  & {\scriptsize Separation}  & {\scriptsize e}  & {\scriptsize i(J2000)}  & {\scriptsize $i_{m}$}  & {\scriptsize $\omega$}  & {\scriptsize D ratio}  & {\scriptsize $D_{p}$}  & {\scriptsize $\Omega$}  & {\scriptsize Ref. }\tabularnewline
 & {\scriptsize Name}  & {\scriptsize (days)}  & {\scriptsize (km)}  &  & {\scriptsize (deg)}  & {\scriptsize (deg)}  & {\scriptsize (deg)}  &  & {\scriptsize (km)}  & {\scriptsize (deg)}  & \tabularnewline
\hline 
{\scriptsize 3749 Balam$^{\star}$}  &  & {\scriptsize 61} {\scriptsize $\pm$10}  & {\scriptsize 289} {\scriptsize $\pm$} {\scriptsize 13}  & {\scriptsize 0.90  } &  &  &  & {\scriptsize 0.43}  &  &  & {\scriptsize 15,25 }\tabularnewline
{\scriptsize 45 Eugenia$^{\star}$}  & {\scriptsize Petit-Prince}  & {\scriptsize 4.77} {\scriptsize $\pm$0.001}  & {\scriptsize 1180} {\scriptsize $\pm$} {\scriptsize 8}  & {\scriptsize 0 } & {\scriptsize 109 $\pm$ 2} & {\scriptsize 126.78}  & {\scriptsize 112} & {\scriptsize 0.81}  & {\scriptsize 202}  & {\scriptsize 203 $\pm$} {\scriptsize 2}  & {\scriptsize 16,17 }\tabularnewline
{\scriptsize 22 Kalliope}  & {\scriptsize Linus}  & {\scriptsize 3.6} {\scriptsize $\pm$0.001}  & {\scriptsize 1095} {\scriptsize $\pm$} {\scriptsize 11}  & {\scriptsize 0 } & {\scriptsize 99.6 $\pm$ 0.5} & {\scriptsize 103.42}  & {\scriptsize -92.5 $\pm$ 60} & {\scriptsize 0.15}  & {\scriptsize 181} {\scriptsize $\pm$4.6}  & {\scriptsize 284.5 $\pm$} {\scriptsize 2}  & {\scriptsize 16,17 }\tabularnewline
{\scriptsize 283 Emma}  &  & {\scriptsize 3.35} {\scriptsize $\pm$0.00093 } & {\scriptsize 581} {\scriptsize $\pm$} {\scriptsize 3.6}  & {\scriptsize 0.12 $\pm$ 0.01} & {\scriptsize 94.2 $\pm$ 0.4} & {\scriptsize 65.46}  & {\scriptsize 40 $\pm$ 4} & {\scriptsize 0.06}  & {\scriptsize 160}  & {\scriptsize 345.4} {\scriptsize $\pm$ 0.4}  & {\scriptsize 16,25 }\tabularnewline
{\scriptsize 130 Elektra}  &  & {\scriptsize 5.26} {\scriptsize $\pm$0.0053}  & {\scriptsize 1318} {\scriptsize $\pm$} {\scriptsize 24}  & {\scriptsize 0.13 $\pm$ 0.03} & {\scriptsize 25 $\pm$ 2} & {\scriptsize 23.75}  & {\scriptsize 311 $\pm$ 5} & {\scriptsize 0.04}  & {\scriptsize 215}  & {\scriptsize 1.6} {\scriptsize $\pm$ 2}  & {\scriptsize 15,25 }\tabularnewline
{\scriptsize 379 Huenna}  &  & {\scriptsize 87.6} {\scriptsize $\pm$0.026}  & {\scriptsize 3335.8} {\scriptsize $\pm$} {\scriptsize 54.9}  & {\scriptsize 0.22 $\pm$ 0.01} & {\scriptsize 152.7 $\pm$ 0.3} & {\scriptsize 168.84}  & {\scriptsize 284 $\pm$ 5} & {\scriptsize 0.06}  & {\scriptsize 92}  & {\scriptsize 204.3} {\scriptsize $\pm$ 0.3}  & {\scriptsize 15,25 }\tabularnewline
{\scriptsize 762 Pulcova}  &  & {\scriptsize 4.44} {\scriptsize $\pm$0.001}  & {\scriptsize 703} {\scriptsize $\pm$} {\scriptsize 13}  & {\scriptsize 0.03 $\pm$ 0.01} & {\scriptsize 132 $\pm$ 2} & {\scriptsize 131.9}  & {\scriptsize -189 $\pm$ 20} & {\scriptsize 0.14}  & {\scriptsize 137} {\scriptsize $\pm$3.2}  & {\scriptsize 235} {\scriptsize $\pm$ 2}  & {\scriptsize 16,25 }\tabularnewline
{\scriptsize 90 Antiope}  &  & {\scriptsize 0.69} {\scriptsize $\pm4.1\times10^{-6}$}  & {\scriptsize 171} {\scriptsize $\pm$} {\scriptsize 13}  & {\scriptsize 0.03 } & {\scriptsize 63.70 $\pm$ 2} & {\scriptsize 54.74}  & {\scriptsize 60 $\pm$ 30} & {\scriptsize 0.95}  & {\scriptsize 87.8} {\scriptsize $\pm$1}  & {\scriptsize 303.1} {\scriptsize $\pm$ 2}  & {\scriptsize 6,17 }\tabularnewline
{\scriptsize 121 Hermione}  &  & {\scriptsize 2.56} {\scriptsize $\pm$0.0021}  & {\scriptsize 747} {\scriptsize $\pm$} {\scriptsize 13}  & {\scriptsize 0 } & {\scriptsize 79.1 $\pm$ 4} & {\scriptsize 70.19}  & {\scriptsize 84.3 } & {\scriptsize 0.17}  & {\scriptsize 187} {\scriptsize $\pm$68}  & {\scriptsize 83.7} {\scriptsize $\pm$ 3}  & {\scriptsize 7,13 }\tabularnewline
{\scriptsize 107 Camilla}  &  & {\scriptsize 3.72} {\scriptsize $\pm$0.003}  & {\scriptsize 1250} {\scriptsize $\pm$} {\scriptsize 10} & {\scriptsize 0 } & {\scriptsize 17 $\pm$ 5} & {\scriptsize 28.32}  & {\scriptsize -32 } & {\scriptsize 0.06}  & {\scriptsize 249}  & {\scriptsize 141.00} {\scriptsize $\pm$ 2}  & {\scriptsize 15,25 }\tabularnewline
{\scriptsize 87 Sylvia$^{\star}$}  & {\scriptsize Romulus}  & {\scriptsize 3.65} {\scriptsize $\pm$0.0007}  & {\scriptsize 1356} {\scriptsize $\pm$ 5}  & {\scriptsize 0 } & {\scriptsize 7 } & {\scriptsize 27.89}  & {\scriptsize -87 $\pm$ 11} & {\scriptsize 0.46}  & {\scriptsize 282} {\scriptsize $\pm$4}  & {\scriptsize 101}  & {\scriptsize 13,25 }\tabularnewline
{\scriptsize 617 Patroclus}  & {\scriptsize Menoetius}  & {\scriptsize 4.28} {\scriptsize $\pm$0.004}  & {\scriptsize 680} {\scriptsize $\pm$} {\scriptsize 5}  & {\scriptsize 0.02 $\pm$ 0.02} &  &  &  & {\scriptsize 0.92}  & {\scriptsize 60.9}  &  & {\scriptsize 17 }\tabularnewline
{\scriptsize 42355 Typhon}  & {\scriptsize Echidna}  & {\scriptsize 18.97} {\scriptsize $\pm$0.0064}  & {\scriptsize 1628} {\scriptsize $\pm$} {\scriptsize 5}  & {\scriptsize 0.53 $\pm$ 0.02} & {\scriptsize 37.9 $\pm$ 2} & {\scriptsize 50.56}  & {\scriptsize 99 } & {\scriptsize 0.55}  & {\scriptsize 76} {\scriptsize $_{-16}^{+14}$ } & {\scriptsize 253.1} {\scriptsize $\pm$ 4}  & {\scriptsize 10,21 }\tabularnewline
{\scriptsize 1999OJ4}  &  & {\scriptsize 84.09} {\scriptsize $\pm$0.016}  & {\scriptsize 3303} {\scriptsize $\pm$} {\scriptsize 5}  & {\scriptsize 0.37 $\pm$ 0.01} & {\scriptsize 53.80 $\pm$ 1.2} & {\scriptsize 119.56}  & {\scriptsize 53.96 } &  & {\scriptsize 37.5} {\scriptsize $\pm$8.5}  & {\scriptsize 275.8} {\scriptsize $\pm$ 2.2} & {\scriptsize 11 }\tabularnewline
 &  & {\scriptsize 84.14} {\scriptsize $\pm$0.016}  & {\scriptsize 3225} {\scriptsize $\pm$} {\scriptsize 18}  & {\scriptsize 0.36 $\pm$ 0.01} & {\scriptsize 99.8 $\pm$ 1.5} & {\scriptsize 56.7}  & {\scriptsize 71.7 } &  & {\scriptsize 37.5} {\scriptsize $\pm$8.5}  & {\scriptsize 210.2} {\scriptsize $\pm$ 1.6} & {\scriptsize 11 }\tabularnewline
{\scriptsize 90482 Orcus}  & {\scriptsize Vanth}  & {\scriptsize 9.54} {\scriptsize $\pm$0.0001 } & {\scriptsize 8980} {\scriptsize $\pm$} {\scriptsize 18}  & {\scriptsize 0 } & {\scriptsize 90.2 $\pm$ 0.6} & {\scriptsize 92.13}  & {\scriptsize 0 } & {\scriptsize 0.31}  & {\scriptsize 900}  & {\scriptsize 50} {\scriptsize $\pm$ 0.6} & {\scriptsize 4 }\tabularnewline
 &  & {\scriptsize 9.5392} {\scriptsize $\pm$0.0001 } & {\scriptsize 8985} {\scriptsize $\pm$} {\scriptsize 24} & {\scriptsize 0 } & {\scriptsize 305.8 $\pm$ 0.6} & {\scriptsize 70.02}  & {\scriptsize 0 } & {\scriptsize 0.31}  & {\scriptsize 900}  & {\scriptsize 249.4} {\scriptsize $\pm$ 0.4} & {\scriptsize 4 }\tabularnewline
{\scriptsize 47171  TC36$^{\star}$}  &   & {\scriptsize 1.9068} {\scriptsize $\pm$0.0001 } & {\scriptsize 867} {\scriptsize $\pm$} {\scriptsize 11}  & {\scriptsize 0.101 $\pm$ 0.006 } & {\scriptsize 88.9 $\pm$ 0.6} & {\scriptsize 74.21}  & {\scriptsize 77.7  } & {\scriptsize 0.93}  & {\scriptsize 265$^{+41}_{-35}$}  & {\scriptsize 330} {\scriptsize $\pm$ 1} & {\scriptsize 1 }\tabularnewline
{\scriptsize 134340 Pluto$^{\star}$}  & {\scriptsize Charon}  & {\scriptsize 6.39} {\scriptsize $\pm10^{-6}$}  & {\scriptsize 19571.4} {\scriptsize $\pm$} {\scriptsize 24} & {\scriptsize 0 } & {\scriptsize 96.16 } & {\scriptsize 119.61}  & {\scriptsize 0 } & {\scriptsize 0.49}  & {\scriptsize 2302}  & {\scriptsize 223.05} {\scriptsize $\pm10^{-4}$ } & {\scriptsize 3,19,21,25 }\tabularnewline
{\scriptsize 134860 2000OJ67}  &  & {\scriptsize 22.04} {\scriptsize $\pm$0.004}  & {\scriptsize 2361} {\scriptsize $\pm$} {\scriptsize 36} & {\scriptsize 0.09 $\pm$ 0.02} & {\scriptsize 84.6 $\pm$ 3} & {\scriptsize 85.21}  & {\scriptsize -233.9 } &  & {\scriptsize 69} {\scriptsize $\pm$16}  & {\scriptsize 272.9} {\scriptsize $\pm$ 3.1} & {\scriptsize 11 }\tabularnewline
 &  & {\scriptsize 22.04} {\scriptsize $\pm$0.0036 } & {\scriptsize 2352} {\scriptsize $\pm$} {\scriptsize 35} & {\scriptsize 0.09 $\pm$ 0.02} & {\scriptsize 73.80 $\pm$ 2.9} & {\scriptsize 94.45}  & {\scriptsize 136.8 } &  & {\scriptsize 69} {\scriptsize $\pm$16}  & {\scriptsize 212.2} {\scriptsize $\pm$ 3.3} & {\scriptsize 11 }\tabularnewline
{\scriptsize 2001XR254}  &  & {\scriptsize 125.61} {\scriptsize $\pm$0.12}  & {\scriptsize 9326} {\scriptsize $\pm$} {\scriptsize 75} & {\scriptsize 0.56 } & {\scriptsize 41.07 $\pm$ 0.22} & {\scriptsize 20.27}  & {\scriptsize -94.76 } &  & {\scriptsize 84.5} {\scriptsize $\pm$19.5}  & {\scriptsize 341.16} {\scriptsize $\pm$ 0.33} & {\scriptsize 11 }\tabularnewline
 &  & {\scriptsize 125.61} {\scriptsize $\pm$0.13 } & {\scriptsize 9211} {\scriptsize $\pm$} {\scriptsize 69} & {\scriptsize 0.55 } & {\scriptsize 154.50 $\pm$ 0.22} & {\scriptsize 155.36}  & {\scriptsize -21.88 } &  & {\scriptsize 84.5} {\scriptsize $\pm$19.5}  & {\scriptsize 125.18} {\scriptsize $\pm$ 0.55} & {\scriptsize 11 }\tabularnewline
{\scriptsize 136108 Haumea$^{\star}$}  & {\scriptsize Hi'iaka}  & {\scriptsize 49.13} {\scriptsize $\pm$ 0.03}  & {\scriptsize 49500} {\scriptsize $\pm$} {\scriptsize 69} & {\scriptsize 0.05 } & {\scriptsize 234.8 $\pm$ 0.4} & {\scriptsize 139.14}  & {\scriptsize 278.6 $\pm$ 0.4} &  &  & {\scriptsize 26.1} {\scriptsize $\pm$ 0.4} & {\scriptsize 2,21,24 }\tabularnewline
{\scriptsize 66652 Borasisi}  &  & {\scriptsize 46.26} {\scriptsize $_{-0.065}^{+0.006}$}  & {\scriptsize 4660} {\scriptsize $\pm$} {\scriptsize 170} & {\scriptsize 0.46 $\pm$ 0.01} & {\scriptsize 152 $\pm$ 3} &  & {\scriptsize 159.94 $\pm$ 4.01} & {\scriptsize 1}  & {\scriptsize 316}  &  & {\scriptsize 20,25 }\tabularnewline
 &  & {\scriptsize 46.23} {\scriptsize $_{-0.074}^{+0.006}$}  & {\scriptsize 4700} {\scriptsize $\pm$} {\scriptsize 170} & {\scriptsize 0.45 $\pm$ 0.01} & {\scriptsize 51 $\pm$ 3} &  & {\scriptsize 167.39 $\pm$ 3.44} & {\scriptsize 1}  & {\scriptsize 316}  &  & {\scriptsize 20,25 }\tabularnewline
{\scriptsize 2001QW322}  &  & {\scriptsize 9855}  & {\scriptsize 114000} {\scriptsize $\pm$}  & {\scriptsize 0.2 } & {\scriptsize 118 } &  &  &  & {\scriptsize 54}  &  & {\scriptsize 22 }\tabularnewline
 &  & {\scriptsize 6570}  & {\scriptsize 105000} {\scriptsize $\pm$}  & {\scriptsize 0.4 } & {\scriptsize 130 } &  &  &  & {\scriptsize 58}  &  & {\scriptsize 23 }\tabularnewline
{\scriptsize 88611 Teharonhiawako}  & {\scriptsize Sawiskera}  & {\scriptsize 876} {\scriptsize $\pm$227} & {\scriptsize 31409} {\scriptsize $\pm$} {\scriptsize 2500} & {\scriptsize 0.31 $\pm$ 0.08} & {\scriptsize 128.1 $\pm$6.5} & {\scriptsize 128.8}  & {\scriptsize 330.3 $\pm$ 22.4} & {\scriptsize 0.69}  & {\scriptsize 78}  & {\scriptsize 96.70} {\scriptsize $\pm$ 13.4} & {\scriptsize 22,25 }\tabularnewline
{\scriptsize 2003TJ58}  &  & {\scriptsize 137.32} {\scriptsize $\pm$0.19}  & {\scriptsize 3799} {\scriptsize $\pm$} {\scriptsize 54} & {\scriptsize 0.53 $\pm$ 0.01} & {\scriptsize 38.1 $\pm$ 2.1} & {\scriptsize 62.25}  & {\scriptsize -110.04 } &  & {\scriptsize 32.5} {\scriptsize $\pm$7.5}  & {\scriptsize 194.60} {\scriptsize $\pm$ 4.2} & {\scriptsize 11 }\tabularnewline
 &  & {\scriptsize 137.32} {\scriptsize $\pm$0.19}  & {\scriptsize 3728} {\scriptsize $\pm$} {\scriptsize 44} & {\scriptsize 0.53 $\pm$ 0.01} & {\scriptsize 96.1 $\pm$ 2} & {\scriptsize 116.77}  & {\scriptsize -88.90 } &  & {\scriptsize 32.5} {\scriptsize $\pm$7.5}  & {\scriptsize 150.80} {\scriptsize $\pm$ 2.8} & {\scriptsize 11 }\tabularnewline
{\scriptsize 1998WW31}  &  & {\scriptsize 574} {\scriptsize $\pm$10}  & {\scriptsize 22300} {\scriptsize $\pm$ 44} & {\scriptsize 0.82 $\pm$ 0.05} & {\scriptsize 41.7 $\pm$ 0.7} & {\scriptsize 51.96}  & {\scriptsize 159.50 } & {\scriptsize 0.83}  & {\scriptsize 118}  & {\scriptsize 94.30} {\scriptsize $\pm$ 0.8} & {\scriptsize 25,26 }\tabularnewline
{\scriptsize 2004PB108}  &  & {\scriptsize 97.02} {\scriptsize $\pm$0.07}  & {\scriptsize 10400} {\scriptsize $\pm$} {\scriptsize 130} & {\scriptsize 0.44 $\pm$ 0.01} & {\scriptsize 89 $\pm$ 1.1} & {\scriptsize 84.13}  & {\scriptsize 229.93 } &  & {\scriptsize 120.5} {\scriptsize $\pm$27.5 } & {\scriptsize 121.99} {\scriptsize $\pm$ 0.75} & {\scriptsize 11 }\tabularnewline
 &  & {\scriptsize 97.08} {\scriptsize $\pm$0.069 } & {\scriptsize 10550} {\scriptsize $\pm$} {\scriptsize 130} & {\scriptsize 0.45 $\pm$ 0.01} & {\scriptsize 106.55 $\pm$ 0.99} & {\scriptsize 95.23}  & {\scriptsize 211.91 } &  & {\scriptsize 120.5} {\scriptsize $\pm$27.5}  & {\scriptsize 30.19} {\scriptsize $\pm$ 0.86} & {\scriptsize 11 }\tabularnewline
{\scriptsize 58534 Logos}  & {\scriptsize Zoe}  & {\scriptsize 312} {\scriptsize $\pm$3}  & {\scriptsize 8010} {\scriptsize $\pm$} {\scriptsize 80}  & {\scriptsize 0.45 $\pm$ 0.03} & {\scriptsize 121.5 $\pm$ 2} &  & {\scriptsize 310.13 $\pm$ 2.87} & {\scriptsize 0.825}  & {\scriptsize 80}  &  & {\scriptsize 19,25 }\tabularnewline
 &  & {\scriptsize 310} {\scriptsize $\pm$3}  & {\scriptsize 7970} {\scriptsize $\pm$ 80}  & {\scriptsize 0.37 $\pm$ 0.01} & {\scriptsize 69 $\pm$ 2} &  & {\scriptsize 298.09 $\pm$ 5.73} & {\scriptsize 0.825}  & {\scriptsize 80}  &  & {\scriptsize 19,25 }\tabularnewline
{\scriptsize 2000QL251}  &  & {\scriptsize 56.46} {\scriptsize $\pm$0.018}  & {\scriptsize 4991} {\scriptsize $\pm$} {\scriptsize 17} & {\scriptsize 0.49 $\pm$ 0.01} & {\scriptsize 127.78 $\pm$ 0.62} & {\scriptsize 135.7}  & {\scriptsize 42.2 } &  & {\scriptsize 74} {\scriptsize $\pm$17}  & {\scriptsize 109.5} {\scriptsize $\pm$ 1.1} & {\scriptsize 11 }\tabularnewline
 &  & {\scriptsize 56.44} {\scriptsize $\pm$0.017}  & {\scriptsize 5014} {\scriptsize $\pm$} {\scriptsize 16} & {\scriptsize 0.49 $\pm$ 0.01} & {\scriptsize 45.62 $\pm$ 0.66} & {\scriptsize 46.58}  & {\scriptsize 45.70 } &  & {\scriptsize 74} {\scriptsize $\pm$17}  & {\scriptsize 71.20} {\scriptsize $\pm$ 1.1} & {\scriptsize 11 }\tabularnewline
{\scriptsize 136199 Eris}  & {\scriptsize Dysnomia}  & {\scriptsize 15.77} {\scriptsize $\pm$0.002}  & {\scriptsize 37430} {\scriptsize $\pm$} {\scriptsize 140} & {\scriptsize 0.01 } & {\scriptsize 61.3 $\pm$ 0.7} & {\scriptsize 94.98}  &  &  & {\scriptsize 2400} {\scriptsize $\pm$100}  & {\scriptsize 139} {\scriptsize $\pm$ 1} & {\scriptsize 3,8,21 }\tabularnewline
 &  & {\scriptsize 15.77} {\scriptsize $\pm$0.002 } & {\scriptsize 37370} {\scriptsize $\pm$} {\scriptsize 150}  & {\scriptsize 0.01 } & {\scriptsize 142 $\pm$ 3} & {\scriptsize 85.74}  &  &  & {\scriptsize 2400} {\scriptsize $\pm$101}  & {\scriptsize 68} {\scriptsize $\pm$ 3} & {\scriptsize 3,8,21}\tabularnewline
{\scriptsize 65489 Ceto}  & {\scriptsize Phorcys}  & {\scriptsize 9.55} {\scriptsize $\pm$0.007 } & {\scriptsize 1841} {\scriptsize $\pm$} {\scriptsize 47} & {\scriptsize 0.02 } & {\scriptsize 116.6 $\pm$ 3} & {\scriptsize 115.38}  & {\scriptsize -64.6 } & {\scriptsize 0.17}  & {\scriptsize 87}  & {\scriptsize 134.6} {\scriptsize $\pm$ 3.4} & {\scriptsize 9,23 }\tabularnewline
 &  & {\scriptsize 9.56} {\scriptsize $\pm$0.008}  & {\scriptsize 1840} {\scriptsize $\pm$} {\scriptsize 47} & {\scriptsize 0.01 } & {\scriptsize 68.8 $\pm$ 2.9} & {\scriptsize 66.28}  & {\scriptsize -65.5 } & {\scriptsize 0.79}  & {\scriptsize 87}  & {\scriptsize 105.5} {\scriptsize $\pm$ 3.7} & {\scriptsize 9,23 }\tabularnewline
\hline 
\multicolumn{12}{l}{{\scriptsize References: $^{1}$ \citet{ben+09b}  $^{2}$ \citet{bro+05} $^{3}$ \citet{bro+07}
$^{4}$ \citet{bro+10} $^{5}$ \citet{bui+06} $^{6}$ \citet{des+07}
$^{7}$ \citet{des+09} }}\tabularnewline
\multicolumn{12}{l}{{\scriptsize $^{8}$ \citet{gre+08}$^{9}$ \citet{gru+07} $^{10}$
\citet{gru+08} $^{11}$ \citet{gru+09} $^{12}$ \citet{hes+05}
$^{13}$ \citet{mar+05} }}\tabularnewline
\multicolumn{12}{l}{{\scriptsize $^{14}$ \citet{mar+06} $^{15}$ \citet{mar+08} $^{15}$
\citet{mar+08b} $^{17}$ \citet{marg+01} $^{18}$ \citet{nol+03}
$^{19}$ \citet{nol+04} $^{20}$ \citet{nol+04b}}}\tabularnewline
\multicolumn{12}{l}{{\scriptsize{} $^{21}$ \citet{nol+08b} $^{22}$ \citet{osi+03} $^{23}$\citet{pet+08}
$^{24}$ \citet{rab+06} $^{25}$ \citet{ric+06} $^{26}$ \citet{vei+02}}}\tabularnewline
\hline 
\multicolumn{12}{l}{}\tabularnewline
\end{tabular}

 Errors estimates are also shown where available. Unnamed satellites are omitted. Multiple systems are marked with ($^{\star}$) and the orbital parameters are for the (listed) outer satellite. All measured orbital parameters are given with respect to J2000 equatorial plane. Calculation method of the mutual inclinations is given in the appendix.
\end{sidewaystable*}

\begin{table*}
\caption{The external (helicentric) orbital parameters of BMPs}
\begin{tabular}{lllllllll}
\hline 
Name  & SMA  & $e_{out}$  & $i_{out}$  & $\Omega_{out}$  & Mass  & error  & class  & ref. \tabularnewline
 & (AU)  &  & (deg)  & (deg)  & $(10^{18}$ kg)  & $(10^{18}$ kg)  &  & \tabularnewline
\hline 
3749 Balam$^{\star}$  & 2.24  & 0.11  & 5.39  & 295.84  & 0  & $\pm$0.00002  & FF  & \tabularnewline
45 Eugenia$^{\star}$  & 2.72  & 0.08  & 6.61  & 147.92  & 5.69  & $\pm$0.12  & MB  & \tabularnewline
22 Kalliope  & 2.91  & 0.1  & 13.71  & 66.23  & 8.10  & $\pm$0.2  & MB  & \tabularnewline
283 Emma  & 3.04  & 0.15  & 8  & 304.42  & 1.38  & $\pm$0.03  & EF  & \tabularnewline
130 Elektra  & 3.12  & 0.21  & 22.87  & 145.46  & 6.6  & $\pm$0.4  & MB  & \tabularnewline
379 Huenna  & 3.13  & 0.19  & 1.67  & 172.07  & 0.38  & $\pm$0.019  & TF  & \tabularnewline
762 Pulcova  & 3.15  & 0.1  & 13.09  & 305.8  & 1.4  & $\pm$0.1  & MB  & \tabularnewline
90 Antiope  & 3.16  & 0.16  & 2.22  & 70.22  & 0.83  & $\pm$0.02  & TF  & \tabularnewline
121 Hermione  & 3.44  & 0.14  & 7.6  & 73.18  & 4.7  & $\pm$0.2  & OMB  & \tabularnewline
107 Camilla  & 3.48  & 0.08  & 10.05  & 173.12  & 11.2  & $\pm$0.3  & OMB  & \tabularnewline
87 Sylvia$^{\star}$  & 3.49  & 0.08  & 10.86  & 73.31  & 14.87  & $\pm$0.06  & OMB  & \tabularnewline
617 Patroclus  & 5.22  & 0.14  & 22.05  & 44.35  & 1.36  & $\pm$0.11  & JT  & \tabularnewline
42355 Typhon  & 37.65  & 0.53  & 2.43  & 351.96  & 0.95  & $\pm$0.052  & Cent  & \tabularnewline
1999OJ4  & 38.10  & 0.02  & 2.61  & 127.46  & 0.40  & $\pm$0.0087  & ICC  & 11 \tabularnewline
90482 Orcus  & 39.17  & 0.23  & 20.58  & 268.65  & 632  & $\pm5$  & 3:2N  & 4 \tabularnewline
47171  TC36$^{\star}$& 39.7  & 0.23  & 8.41  & 97.08 & 14.2  & $\pm0.05$  & 3:2N  &  \tabularnewline
134340 Pluto$^{\star}$  & 39.45  & 0.25  & 17.09  & 110.38  & 14570  & $\pm$9  & 3:2N  & \tabularnewline
134860 2000OJ67  & 42.9  & 0.01  & 1.33  & 96.76  & 2.15  & $\pm$0.099  & CC  & 11 \tabularnewline
2001 XR254  & 43  & 0.02  & 2.66  & 52.73  & 3.92  & $\pm$0.089  & CC  & 11 \tabularnewline
136108 Haumea$\star$  & 43.08  & 0.2  & 28.22  & 122.1  & 4200  & $\pm$100  & HF  & \tabularnewline
66652 Borasisi  & 44.07  & 0.09  & 0.56  & 84.74  & 3.8  & $\pm$0.4  & CC  & \tabularnewline
2001QW322  & 44.28  & 0.02  & 4.8  & 124.67  & 1.5  &  & CC  & \tabularnewline
88611 Teharonhiawako  & 44.29  & 0.02  & 2.57  & 304.63  & 3.2  & $_{-0.2}^{+0.3}$  & CC  & \tabularnewline
2003TJ58  & 44.5  & 0.09  & 1.31  & 37.12  & 0.22  & $\pm$0.0078  & CC  & 11 \tabularnewline
1998WW31  & 44.64  & 0.09  & 6.81  & 237.1  & 2.7  &  & CC  & \tabularnewline
2004PB108  & 45.1  & 0.11  & 19.19  & 147.38  & 9.88  & $\pm$0.37  & HC  & 10 \tabularnewline
58534 Logos  & 45.5  & 0.12  & 2.9  & 132.64  & 0.42  & $\pm$0.02  & CC  & \tabularnewline
2000QL251  & 47.8  & 0.21  & 5.83  & 223.29  & 3.14  & $\pm$0.03  & 2:1N  & 10 \tabularnewline
136199 Eris  & 67.96  & 0.44  & 43.97  & 35.99  & 16600  & $\pm$200  & SD  & \tabularnewline
65489 Ceto  & 100.17  & 0.82  & 22.32  & 172.04  & 5.42  & $\pm$0.42  & Cent  & \tabularnewline
\hline
\end{tabular}

Unless noted otherwise all
outer parameters are taken from JPL small bodies database (see http://ssd.jpl.nasa.gov/sbdb.cgi).
Errors are shown where available. Multiple systems are marked with
$^{\star}$. References are detailed in Table 1. The heliocentric
orbital classification is noted as \textquotedbl{}class\textquotedbl{},
where we used the following notations: MB, OMB, JT, CC, ICC, HC, SD,
Cen, HF, TF, EF, FF and $n:m$N for Main Belt, Outer Main Belt, Jupiter
Trojan, Cold Classical, Inner Cold Classical, Hot Classical, Scattered
Disk, Centaur,Haumea Family, Themis Family, Eos Family, Flora Family
for $n:m$ Neptune resonances respectably. Asteroid family membership
is based on listings in \citet{zap+95}. The prevalence of binaries
among the cold classical population of the transneptunian belt is
discussed further in \citet{nol+08b}.
\end{table*}

Another important bias is that systems with low inclinations%
\footnote{Here, inclination means the mutual inclination between the heliocentric
orbit and the mutual binary orbit; for objects at these great distances,
the difference between heliocentric and geocentric viewing angles
is not significant for this bias.%
} present nearly edge-on orbits with respect to Earth-based
observations, while systems with high relative inclinations are usually
seen face-on. Since binaries are discovered when the components are
significantly separated on-the-sky, there is a greater likelihood
for low inclination secondaries to be unresolvable when observed at
a single random epoch. This bias is reduced as the projected semi-major
axis grows, but remains significant even at a few times the resolution
limit.

\subsubsection{Binary asteroids}

While some binary asteroids have been imaged by HST, ground-based
adaptive optics has been employed more often for these brighter systems
than their trans-Neptunian counterparts. The much smaller sizes of
typical asteroids is partly offset by their increased brightness and
proximity. Discovery of binaries photometrically through double-periodic
light curves and/or mutual events is common for near-Earth asteroids
\citep[][ note, however that these are not included in our analysis, and are mentioned here for completeness]{Pra+06}.
Other methods are radar observations that often reveal near-Earth
binary asteroids (not discussed here), and stellar occultations that
can reveal the presence of main belt binary asteroids \citep[e.g. ][]{des+07}.
Theoretically, the radar and occultations techniques have fewer 
observational biases than the other more common techniques, 
but their application is severely
limited.

Binary asteroids are also clearly subject to detection bias: objects
with smaller satellites are more difficult to observe. However, since
asteroids are searched for binaries using ground-based facilities,
there is a greater possibility of searching for companions in more
than a single snapshot. Furthermore, the geocentric orientations of
these systems change much more rapidly than for essentially fixed
KBO orbits ($\sim60^\circ\,yr^{-1}$  for asteroids, vs. $\sim2^\circ\,yr^{-1}$ for KBOs).
 Therefore, the eccentricity and inclination biases are
not as strong as in the Kuiper belt. In binaries discovered through
mutual events, there is an obvious bias towards edge-on systems, though
\textquotedbl{}edge-on\textquotedbl{} can probe a wide range of inclinations.

\section{Discussion}

Several different processes affect the mutual orbits of
BMPs. Some are related to their initial formation and others to their
later evolution either as isolated systems, or due to the effects of
external perturbations and encounters with other objects. The
studies of these processes have been focused on a specific type of BMPs such 
as binary asteroids or binary TNOs. However, all of these suggested processes
could in principle be relevant for the formation/evolution of BMPs both close (binary asteroids) and far (binary TNOs) from the sun. 

The various suggested mechanisms for the formation of BMPs 
\citep[see refs. ][for some overviews]{ast+05,ric+06,nol+08b} predict different
initial orbital configurations. These include the following:
\begin{itemize}
\item Smashed target satellites \citep[SMATS; ][]{wei+02,dur+04,dur+10}: Low eccentricity distribution expected from collisionally formed satellites orbiting 
the main collision remnant body (which typically form as close
binaries and are likely to be affected by tides).
\item Escape ejecta binaries \citep[EEBs; ][]{dur+04,dur+10}: Intermediate eccentricities from bound ejecta pairs ejected following a collision of two larger bodies.
\item Exchanged binaries \citep{fun+04}: Typically high eccentricities ($>0.8$) for high mass ratio binaries formed through exchanges.

\item Dynamical friction and chaos assisted capture (CAC) binaries \citep{gol+02,lee+07}: Intermediate eccentricities ($0.2<e<0.8$) for binaries formed through chaos assisted capture of satellites.
\end{itemize} 
Unfortunately, only a few studies explored aspects of the inclination
distribution of BMPs \citep{ast+05,naz+07,sch+08,per+09}.

After their formation, BMPs can be affected by several processes which
can change their orbits. Tidal effect are most important when the
BMPs components approach each other at a close distance. These effects
couple the orbital evolution of the BMP to the spin of the BMPs components,
and the tides raised on the objects serve to dissipate the total angular
momentum of the system. Tidal effects can also excite and enlarge
the eccentricity and inclination of a given BMP \citep[e.g. ][]{gol+66}.
At long enough timescales (depending on specific configuration), 
however, tidally evolved systems are expected
to relax into more circularized configuration, possibly locked configurations
and even mergers. Such effects are thought to produce the period-eccentricity
distributions of (close) binary systems such as stellar binaries and planetary
systems, and are likely to play a similar role in BMP systems \citep{maz08}.

Another evolutionary process is the Kozai-Lidov \citep{koz62,lid62}
mechanism, which is the effect of a secular perturbation from a third
object (in a triple system, i.e. the Sun serves as the third companion
for BMPs) on the (bound) binary system. It could lead to large (order
unity) periodic oscillations (Kozai cycles) in the eccentricity and
inclination, i.e. it could both raise and lower the inclinations and
eccentricities of a system. Note, however, that such a
process is effective only for systems with initially 
high inclinations (($40^{\circ}>i_{m}<140^{\circ}$; with a somewhat
wider inclination range for initially eccentric systems).

The combined effects of the Kozai mechanism in addition to tidal friction,
\citep[Kozai cycles and tidal friction;
KCTF][]{maz+79,kis+98}, can change the orbital parameters of the BMPs, and reduce both the
eccentricity and separation of the BMPs \citep{per+09}. In essence
this effect rapidly lowers the eccentricity of BMPs and shortens their
period.

These mechanisms are effective in isolated systems (although including
the Sun). In collisional systems, encounters between BMPs and other
minor planets or BMPs can change the orbital parameters of the BMPs.
The distribution of inclinations in such systems is likely to be randomized,
where as the eccentricity distribution is expected to approach high
eccentricities, on average \citep{fun+04}, possibly producing a thermal
like distribution \citep{heg75}.

In the following we discuss the implications of the observed distributions
in the light of the dynamical processes discussed above.

\subsection{Eccentricities}

\begin{figure}
\centering \includegraphics[width=84mm]{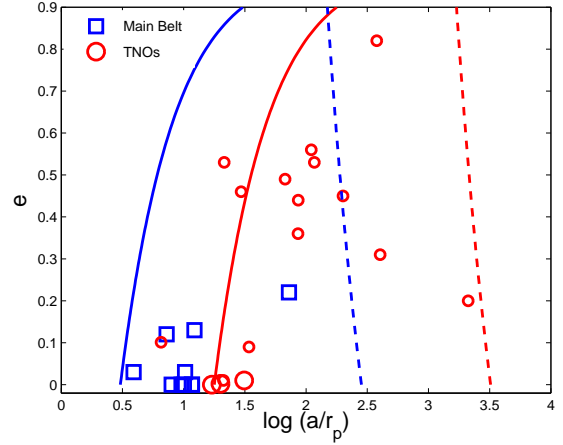} 
\caption{Eccentricity as a function of the binary separations
  normalized to the primary's radius. We consider both TNOs (open
  circles) and asteroids (squares). The large circles correspond to the largest size primary TNOs ($> 500\,km$). The solid line shows the critical separation-eccentricity beyond which typical binaries are expected to be strongly affected by tides and be circularized  (where we choose Pulcova and OJ4 as representative for the asteroid and TNO binaries respectively). The dotted lines show the the critical separation-eccentricity beyond which typical binaries become unstable due to the perturbation by the Sun (the Hill radius); lines are shown for both asteroids (left line) and TNOs (right), as the Hill radius is strongly dependent on the distance from the Sun..}
\label{fig:eVsa_rp}
\end{figure}

The observed eccentricity distribution of BMPs (Figs \ref{fig:histe})
shows both low (and zero) eccentricity BMPs as well as high eccentricity
ones (up to 0.82 and 0.22 for binary TNOs and asteroids, respectively).
The clear correlation between the eccentricity and the semi-major axis of
the BMPs (eccentricity-period distribution of BMPs is shown in fig~\ref{fig:eVsa_rp}), reminiscent of other binary populations
\citep[e.g. binary stars, exoplanets; ][]{maz08} indicates that low eccentricity BMPs
are likely to be produced by tidal circularization, which become important
for BMPs of small separations. In principle, the period eccentricity
distribution can be used to constrain tidal evolution theories and/or
the physical parameters of BMPs which affect the tidal evolution \citep[e.g.
the Q parameter, and its evolution][]{efr+07}. Although current
statistics are still too small to produce strong constraints on such
theories/parameters, one can already check specific tidal evolution
cases (see e. g. the theoretical lines shown in fig. \ref{fig:eVsa_rp}).

The apparent lack of zero and low ($<0.2$) eccentricity binary TNOs at larger
separations (where tidal forces are not effective) suggests that either
the formation processes of binary TNOs are not inclined to form them
at such eccentricities, e.g. in the EEB and CAC scenarios for BMPs formation, and/or that later dynamical evolution changed
their eccentricities. The current sample of binary TNOs, showing the lack
of many high eccentricity BMPs ($e>0.8$), is already large enough
to rule out the exchange formation scenario for binary TNOs as formulated
by \citet{fun+04} as the main single process producing their current
distribution. We note that the lack of high eccentricity for 
the largest binary TNOs together with the frequent intermediate 
eccentricity population of the lower mass binary TNOs could be consistent 
with a collisional scenario. In this process EEBs are made from 
smaller remnant bodies following a collision where as
 the bigger remnants may form SMATs with lower eccentricities 
in non-disruptive collsions \citep{dur+04}.  

The eccentricity distribution of binary asteroids appears to differ from that of binary TNOs, with a larger fraction of binaries at shorter more circular orbits.
However the general trend is similar to that of binary TNOs, showing 
close binaries to typically have more circular orbits 
and wider ones to have higher eccentricities.  Such distribution 
could be consistent with that of binary TNOs, 
given the selection effects (e.g. the difficulty in finding close small sized binary TNOs) and the small statistics. Moreover, the lack of wider period binaries
with higher eccentricities is likely related to the much smaller phase space available to binary asteroids, due to their smalle Hill radii (see fig. 3).
The binary asteroids distribution may therefore also be 
suggestive of a collisional origin. The known collisional families in the 
asteroid belt give further support for this scenario.   

\subsection{Inclinations}

The distribution of BMPs inclinations shows a large fraction of them
to have high inclinations. The inclinations of binary asteroids are
consistent with a random distribution of inclinations (flat in $\cos i$).
The underlying true inclination distribution of the larger sample
of binary TNOs can not be derived directly (given the degenerate inclinations
solutions for most of the sample). Nevertheless, we can statistically
verify whether it too could be consistent with a flat distribution.
To do so, we consider all the possible inclination distributions of
the binary TNOa (i.e. $2^{N_{deg}}$, where $N_{deg}=9$ is the number
of orbits with two degenerate solutions used). We then use the Two-sample
Kolmogorov-Smirnov test to check whether each possible distribution
is consistent with it being drawn from a flat distribution (in $\cos{i}$).
We find that  all of  these distributions are  consistent  
with an isotropic distribution.
We conclude that
the inclinations of both binary TNOs and asteroids are consistent
with a random distribution of inclinations (flat in $\cos i$); clearly
BMPs are not restricted to planar configurations as suggested by \citet{gol+02} and \citet{sch+08}.
In some scenarios \citep{gol+02}, BMPs are formed in a thin planetesimals disk 
with low velocity dispersion. It was suggested that BMPs with high 
inclinations are therefore not likely to be produced under these conditions. 
However, three body encounters can
easily change the inclinations of BMPs, as these could be highly chaotic,
and produce highly inclined orbits even under such conditions (Perets
\& Kupi, in prep.). Collisoinaly formed BMPs are also likely to produce a range of inclination, as material could be ejected in a wide fan, and specifcially EEB could have random inclinations (however further studies in this direction are required). The high inclinations of BMPs could therefore be suggestive
of the collisional environment at which BMPs were formed. Note that since 
encounters between BMPs and other single planetesimal 
can easily erase the initial distribution
of BMPs inclinations, predictions of the inclination distribution
such as suggested by \citet{sch+08}, which do not seem to be consistent
with the currently observed distribution of inclinations (i.e. non
planar configurations), are not likely to constrain formation scenarios
of BMPs.

We note that the combined processes of secular Kozai evolution (due
to perturbations by the sun%
\footnote{Note that alternatively/in addition planets or even a third companion
in triple minor planets systems could also produce such perturbations.%
}) and tidal friction (KCTF) can lead to specific correlations between
inclinations and separations of BMPs (as well as eccentricities).
These processes and the currently observed relations they may have
produced, have been discussed in detail elsewhere \citep[][see also Ragozzine \& Brown, in prep.]{per+09}. We find
that the probability for the eccentricity and inclinations ($|\cos i|$)
distributions to be uncorrelated is $\sim0.04$ (with the correlation
coefficient found to be $0.7$).

We conclude that although understanding the shape of the inclinations
distribution requires more data, it is clear that high inclinations
serve as the rule and not the exception, and point to an excitation 
mechanism, possibly a highly collisional
environment at the epoch of BMPs formation and/or evolution, with
further KCTF evolution playing a role and producing correlations between
the BMPs orbital properties (separations/eccentricities/inclinations).

\subsection{Dependence on mass/size }

The masses and sizes of BMP components, mostly the smaller secondaries, 
are not accurately known. Nevertheless, some interesting trends 
with masses/sizes may already be observed in our binary TNOs sample. 
Binary TNOs with the largest primaries ($>500\,km$; see table 1 and
 fig. 3), are observed to have satellites at small separations and 
low eccentricities, possibly indicating a collisional 
formation mechanism \citep{can05,bro08,rag+09}.
We also note that these binaries seem to cluster at relatively
 high inclinations (not shown, but see table 1), 
although this may not be statistically
significant given our currently small sample, it may suggest KCTF
(highly efficient at high inclinations) was involved in catalyzing
mergers/collisions or tidal evolution of pre-formed binaries  \citep[see ][]{per+09}.  
We do not see any trends in the current sample of binary asteroids, but note 
that no massive ($>500\,km$) binary asteroids are known. Nevertheless,    
it is worth pointing out that most of the binary asteroids 
share similarities with the massive binary TNOs, 
i.e. small separations and low eccentricities, as discussed above, suggesting a 
similar, possibly collisional, origin.     

\subsection{Binary trans-Neptunian objects vs. binary asteroids}

We find that both the observed populations of binary asteroids and binary TNOs 
present similar orbital properties (e.g. high inclinations).
This suggest that some basic features in their formation and evolution
were similar (e.g. dense collisional environment). Although binary
TNOs show typically much higher eccentricities, most of these binaries
have much wider orbits than those of binary asteroids (both due to
observational selection effects, as well as the much
smaller Hill radii in which binary asteroids can exist; see figure
3). The differences in eccentricities may therefore only reflect the
tendency of closer binaries to be more circularized (since tidal friction 
effects become stronger ). 

\subsection{Multiple systems}

The last few years have seen the discovery and characterization of
the first asteroids and TNOs with multiple satellites, which deserve
special mention. Such systems with well-known published orbits include
Pluto \citep{tho+08}, Haumea \citep{rag+09}, 1999 TC36 \citep{ben+09b}
and Sylvia \citep{mar+06}, while the asteroids Eugenia, Balam, Kleopatra,
Minerva, 2001 SN263, and 1994 CC have only recently been announced as triple
systems. Though the detection biases described in Section 2 are present
for these systems, the majority of these systems were first known
as binaries with additional companions found during subsequent study.
It is therefore difficult to estimate the frequency of multiple systems
in the various populations. Nevertheless, the existence of several
such objects indicates that these are not rare. Detailed observations
of these systems can yield mass determinations for each of the bodies
independently, which is not possible for binaries \citep{tho+08,rag+09}.

Multiple systems provide unique additional leverage in determining
the formation and evolution of binaries. For example, the coplanar
nature of the satellite system of Pluto likely requires a dense collisional
formation \citep{ste+06}, although the detailed formation and evolution
of this system is still not understood \citep{war+06,lit+08}. Even
for systems with unknown orbits, the small sizes and compact configurations
suggest that all of these systems are collisionally formed, except
for 1999 TC36; though multiple episodes of YORP-induced fission may
be relevant for the smaller bodies \citep{wal+08}. The hierarchical
and nearly-equal mass nature of the 1999 TC36 triple system cannot
be explained by a single collision and is likely the result of sequential
formation by capture \citep[e.g. ][]{gol+02,lee+07}. Balam has a very
unusual satellite system, including an outer satellite with a putative
eccentricity of $\sim$0.9 \citep{mar+08}, and possibly an unbound
satellite that separated from Balam less than a million years ago
\citep{vok09}.

These systems also present examples of unique orbital evolution. As
pointed out by \citep{rag+09}, the combination of rapid orbital expansion
(compared to the weak expansion around giant planets, which are ineffective
at dissipating tidal energy) and gravitationally interacting satellites,
creates a unique brand of tidal evolution. Multiple resonance crossings
can excite eccentricities and inclinations, perhaps leading to instability
\citep[e.g., ][]{can+99}. In hierarchical systems with significant mutual
inclination, the Kozai-Lidov effect may destabilize the system \citep{per+09}.
For all multiple systems, survival to the present epoch can be difficult,
and multiples must have been more common in the primordial population.

Additional study of these multiple minor planet systems will provide
unique insights into the formation and evolution of these systems.
See \citet{ben+09b} for an additional review of multiple minor planet
systems, including a table of properties.

\section{Summary}
In this paper we compiled a catalog of BMPS, both binary TNOs and binary asteroids, with full orbital solutions.  We presented a first analysis
of the eccentricity and inclinations distribution of BMPs as well as
their semi-major axis-eccentricity distribution.  This data and its
analysis can be used to study and constrain formation and evolutionary
scenario of BMPs.  Specifically we find high relative inclinations for
the BMPs as well as typically large (but not extremely high)
eccentricities for BMPs not affected by tidal evolution. By themselves
these results already suggest BMPs evolved in dense environment in
which collisions and close gravitational encounters formed/perturbed 
the binaries and strongly affected their orbital evolution.  
We suggest that these
encounters together with secular Kozai evolution and tidal effects,
could have erased much of the direct signatures of the initial formation
of BMPs, as possibly reflected by their observed orbits
 \citep[see also ][]{per+09}.  More theoretical work, however, is required to
understand the different parts played by the initial formation and
configurations of BMPs vs.  their later dynamical evolution.
Especially important are better theoretical predictions for the
observational signatures of different BMP formation scenarios, which
could be compared with the data presented in this paper and
additional future data.
\section*{Acknowledgments}
We thank Eran Ofek for useful discussion regarding the transformation from the ecliptic 
plane to the equatorial frame of reference.
SN acknowledge the partial support by Israel Science Foundation
grant 629/05 and U.S.-Israel Binational Science Foundation grant
2004386. SN and HBP acknowledge the generous support of the Dan
David Fellowship. HBP is a CfA, Rothschild, FIRST and Fullbright fellow. 
HBP also acknowledges the Israeli commercial and industrial
club for their support through the Ilan-Ramon-Fullbright
fellowship.
 DR is grateful for the
support by NASA Headquarters under the Earth and Space Sciences
Fellowship.

\bibliographystyle{apj}

\bibliography{planet-formation}

\appendix
\section{Calculation of the mutual inclination}

The  mutual inclination $i_m$ represent the angle between the inner and the outer orbit.
All of the heliocentric parameters   are given with respect to the ecliptic plane, while the inner orbital parameters are given with respect to the equatorial frame of reference.
Thus, we first transform the heliocentric inclination and  the longitude of ascending node from the ecliptic plane ($i_{c},\Omega_{c}$) to the equatorial frame of reference ($i_{q},\Omega_{q}$). 
Let us define the pole vector with respect to the ecliptic (equatorial) frame, ${\bf P}_c$ (${\bf P}_q$):
\begin{equation}
{\bf P}_{c,q} \left( \begin{array}{ccc}
x\\
y\\ 
z
\end{array} \right)
= \left( \begin{array}{ccc}
\sin\Omega_{c,q}\sin i_{c,q} \\
 -\cos\Omega_{c,q}\sin i_{c,q}\\
 \cos i_{c,q}
 \end{array} \right) \ .
\end{equation}
The rotation matrix from the ecliptic to the equatorial is rotation with respect to the $x$ axis of the invariable plane, i.e.,
\begin{equation}
R(x)_{c\to q}= \left( \begin{array}{ccc}
1 & 0 & 0\\
 0 & \cos\epsilon & -\sin\epsilon \\
0 &  \sin\epsilon & \cos\epsilon \end{array} \right) \ .
\end{equation}
where $\epsilon$ is the ecliptic angle.
Thus, the resulting equations of the transformation from the  ecliptic to the equatorial are:
\begin{eqnarray}
\cos(i_q)&=&\cos(i_c)\cos(\epsilon)-\sin(i_c)\sin(\epsilon)\cos(\Omega_c) \ ,\\ \nonumber
\cos( \Omega_q )\sin(i_q)&=&\cos(i_c)\sin(\epsilon)+\sin(i_c)\cos(\epsilon)\cos(\Omega_c) \ ,\\ \nonumber
\sin(\Omega_q)\sin(i_q)&=&\sin(\Omega_c)\sin(i_c) \ ,
\end{eqnarray}
where from the last two equations we can find $\Omega_q$ using the $atan2(x,y)$ function.

Now we turn to  calculate the mutual inclination. We take the longitude of ascending node and
inclinations with respect to equatorial frame (resulting from the above transformation) of the inner and outer orbit, i.e.,
$i_{\out},\Omega_{\out}$, and $i_{s},\Omega_{s}$, and calculate the
mutual inclinations. Again let us define  the binary outer orbit and inner pole vectors as:
\begin{equation}
{\bf P}_{\out,s}= \left( \begin{array}{ccc}
\sin\Omega_{\out,\in}\sin i_{\out,\in} \\
 -\cos\Omega_{\out,\in}\sin i_{\out,\in}\\
 \cos i_{\out,\in}
 \end{array} \right) \ ,
\end{equation}
and thus, the mutual inclination is simply given by
\begin{equation}
\label{mu_i}
i_{m}=\cos^{-1}\left({\bf P}_{\out}\cdot {\bf P}_{\in}\right) \ .
\end{equation}

\bibliographystyle{apj}

\end{document}